\documentclass[preprint]{jfm}
\usepackage{graphicx}
\usepackage{epsfig}
\usepackage{bm}
\usepackage{amsmath}
\usepackage{amssymb}
\usepackage{wasysym}
\usepackage{epstopdf}
\usepackage{color}
\usepackage{xcolor}
\usepackage{epstopdf}
\usepackage{upgreek}
\usepackage{subfloat}
\usepackage{graphicx}% Include figure files
\usepackage{dcolumn}% Align table columns on decimal point
\usepackage{bm}% bold math
\usepackage{latexsym}
\usepackage{subfloat}
\usepackage[amssymb]{SIunits}
\usepackage{subfig}
\usepackage{psfrag}
\usepackage{xcolor}
\usepackage{natbib}
\usepackage{epstopdf}
\usepackage{amsmath}
\usepackage{comment}
\graphicspath{{Figures//}}
\newcommand{\We}{\ensuremath{\mathrm{We}}}   %Eo

\usepackage[colorlinks=false, bookmarks=true]{hyperref}

\title[The clustering morphology of freely rising deformable bubbles]{The clustering morphology of freely rising deformable bubbles} 
\author[Tagawa, Roghair, Prakash, van Sint Annaland, Kuipers, Sun, and Lohse]
{Yoshiyuki\ns Tagawa$^{1}$,\ns  
Ivo\ns Roghair$^{2}$,\ns 
Vivek\ns  N.\ns Prakash$^{1}$,\ns \break
Martin\ns  van\ns Sint\ns Annaland$^{2}$,\ns 
Hans\ns Kuipers$^{2}$,\ns Chao\ns Sun$^{1}$\ns and Detlef\ns Lohse$^{1}$\ns}

\affiliation{$^1$Physics of Fluids Group, Faculty of Science and Technology, J.M. Burgers Center for Fluid Dynamics, University of Twente, PO Box 217, 7500 AE  Enschede, The Netherlands\\[\affilskip]
$^2$ Multiphase Reactors Group, Department of Chemical Engineering and Chemistry, J.M. Burgers Center for Fluid Dynamics, P.O Box 513, 5600 MB Eindhoven University of Technology, Eindhoven, The Netherlands\\[\affilskip]
}
\date{\today}

\begin{document}
\maketitle

\begin{abstract}  
We investigate the clustering morphology of a swarm of freely rising deformable bubbles. 
A three-dimensional Vorono\"i analysis enables us to quantitatively distinguish between two typical clustering configurations: preferential clustering and a grid-like structure.
The bubble data is obtained from direct numerical simulations (DNS) using the front-tracking method. 
It is found that the bubble deformation, represented by the aspect ratio $\chi$, plays a significant role in determining which type of clustering is realized:
Nearly spherical bubbles with $\chi \lesssim$ 1.015 form a grid-like structure, while more deformed bubbles show preferential clustering.
Remarkably, this criteria for the clustering morphology holds for different diameters of the bubbles, surface tension, and viscosity of the liquid in the studied parameter regime.
The mechanism of this clustering behavior is connected to the amount of vorticity generated at the bubble surfaces.
%Spherical bubbles form a grid-like structure due to less vorticity  generation. The preferential clustering for deformed bubbles is a result of the low pressure region in their wakes, which attract other bubbles.
\end{abstract}

%\textbf{TO DO LIST}
%\begin{itemize}
%%\item Instead of $\alpha$, use bubble radius $R$ over box size $L$, i.e. $R/L$.
%%$\item Figure~\ref{fig:NumberEffect}, use red marker at N = 16
%\item Instead of FIgure~\ref{fig:AspectNormSigma}(b), make three-dimensional figure with $\chi, R/L$, and $\alpha$.
%\subitem Make $\chi$-$\sigma$ figures for smaller $\alpha$ and for larger $\alpha$
%\subitem Expand the region $\chi \sim$ 1 
%\end{itemize}

%%%----------------------------------------INTRODUCTION ------------------------------------------------%%%
%%%----------------------------------------INTRODUCTION ------------------------------------------------%%%
%%%----------------------------------------INTRODUCTION ------------------------------------------------%%%
\section{Introduction}
\label{sec:Introduction}
%The spatial distribution of bubbles rising in a liquid
%Bubbles usually have a finite-size.
%In bubbly flow systems, the bubbles are often distributed inhomogeneously, leading to clustering or preferential concentration. 
Particles dispersed in a flow can distribute inhomogeneously, showing clustering or preferential concentration behavior. This is attributed to the interaction between the two phases, and the inertia of the particles (\cite{Calzavarini2008c,Toschi2009}). 
%Moreover, the bubble deformation characteristics greatly affect the clustering behavior, which has not been investigated in detail. 
%\textcolor{red}{The text below talks about pseudo-turbulence in general, and summarizes results of Martinez al JFM 2010, and Ivo, Julian et al  IJMF 2011} \\ 
A swarm of bubbles rising in a quiescent liquid is a subset of the general case of particles dispersed in a complex flow.
This topic of bubbly flow has applications in bubble columns that are important in the chemical industry, in chemical processes such as oxidation, chlorination, in water treatment, and in the steel industry (\cite{Ullmann2010}).
Freely rising bubbles in an originally still liquid are known to induce liquid velocity fluctuations which result in the so-called ``pseudo-turbulence''. 
Bubble clustering in pseudo-turbulence has attracted much attention because of its importance in applications, and lack of understanding of the fundamental physics (\cite{Zenit2001, Riboux2010, Martinez2010, Roghair2011b}).
%The exact reason is still unknown.
%The clustering results did not agree with each other \cite{Roghair2011}. 
% (in one sense, we see this diagonal tendency here also right?). 
%\textcolor{red}{The text below are adapted from Julian et al 2010 JFM (page 2), and citations etc need to be re-checked :} \\ 
\cite{Bunner2002, Bunner2003} have conducted numerical simulations and found that deformability of the bubbles plays an important role in the clustering phenomenon: 
bubbles with small deformability (spherical bubbles) show a horizontal alignment, while deformed bubbles display a preferential clustering in the vertical direction. 
%  accompanied by the formation of rafts. 
%and numerical simulations \cite{Roghair2011} 
Meanwhile, experiments have found both horizontal and vertical clustering depending on parameters like  bubble deformation, size, and other flow properties (\cite{Zenit2001, Cartellier2001, Martinez2010}).
% studies have also found a mild horizontal clustering 
%Theoretical work \cite{Wijngaarden1993, Wijngaarden2005} using the potential flow assumption, has predicted that rising spherical bubbles will form horizontally aligned rafts.
%\textcolor{red}{(The particle-based Re? It might be needed for comparison with others)    }
%ring of bubbles was studied using the radial pair correlation method, which showed different clustering behavior in the experiments and in the numerics.

In the present work, we revisit the issue of bubble clustering, using a Vorono\"{i} analysis technique, which has been proven to be a powerful tool for quantifying the clustering behavior of bubbles and particles in fluid flow, see e.g. \cite{Monchaux2010, Tagawa2012}, or \cite{Fiabane2012}. 
%A proper quantification of the clustering is crucial to study bubble and particle clustering. 
%Previous studies used various methods, for example segregation indicators \cite{Calzavarini2008a}, box-counting method \cite{Fessler1994, Aliseda2002}, and pair correlation functions \cite{Chen2006,Saw2008},
%Tools like single-point statistical analysis \cite{Calzavarini2008b}, and Kaplan-Yorke dimension \cite{Bec2006,Calzavarini2008c} have been able to obtain global information on particle clustering.
% \citet{Monchaux2010} and \citet{Tagawa2012} have demonstrated the power of Vorono\"{\i} analysis to investigate the global clustering of the particles.
%The quantification of the clustering and the Lagrangian statistics of the clusters have been studied in both two-dimensions \cite{Monchaux2010} and three-dimensions \cite{Tagawa2012}. 
Here, we extend the Vorono\"i analysis to study the geometric morphology of the clusters formed by freely rising deformable bubbles. 
The bubble data are obtained from direct numerical simulations of a swarm of rising bubbles.
\vspace{-.5 cm}
\section{Vorono\"{\i} analysis for clustering morphology of bubbles }
\label{sec:Method}
%\subsection{Vorono\"{\i} analysis}
\label{subsec:Voronoi}

\begin{figure}
	\centering
\includegraphics[width=1\textwidth]{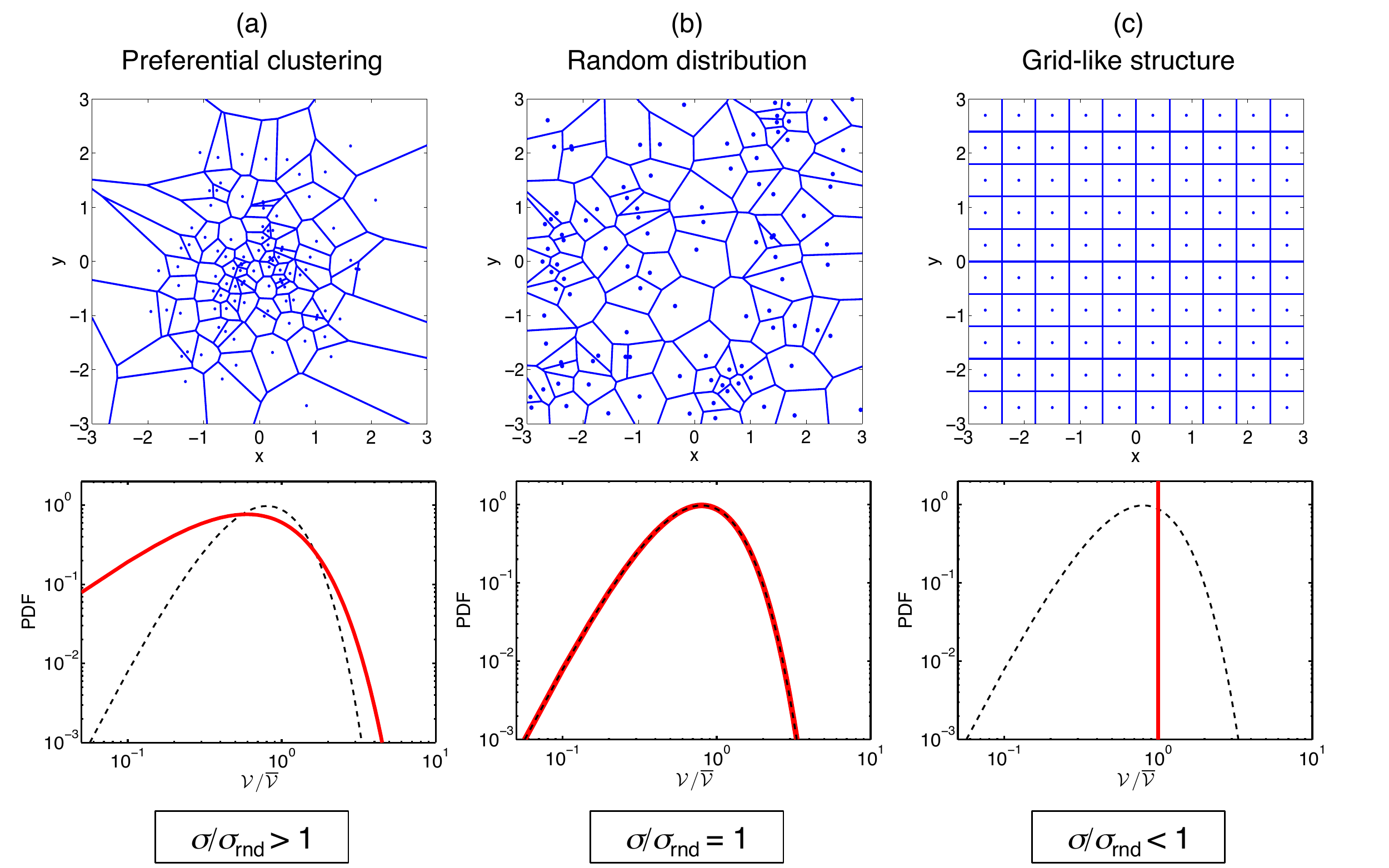}
	\caption{Examples of the different types by which a fixed number (in this case 100) of particles can be spatially distributed : (a) Preferential clustering, (b) randomly distributed, and (c) grid-like structure. The upper figures show the Vorono\"{i} tesselations based on the particle positions in two dimensions, for ease of illustration. The lower figures show the corresponding Probability Density Functions (PDFs) of the Voronoi Volumes (the 3D case). The PDF corresponding to the upper figures (thick red line) are compared with the PDF of the randomly distributed particles (dashed black line). The value of the clustering indicator (i.e. the standard deviation of the PDFs normalized by that of randomly distributed particles, $\sigma$/$\sigma_{rnd}$) are also shown below the PDFs.} %in three cases presented above are : (a) $\sigma$/$\sigma_{rnd} >$ 1, (b) $\sigma$/$\sigma_{rnd}$ = 1, and (c) $\sigma$/$\sigma_{rnd} <$ 1, respectively.}}
	\label{fig:VoronoiExamp}
	\vspace{-.5 cm}
\end{figure}

In the method of Vorono\"{\i}  tesselations, each Vorono\"{\i} cell is defined at a particle location based on its neighbors %ing particles.
(\cite{SpatialTesselation}). 
Every point inside a Vorono\"{\i} cell is the nearest to the particle location compared to the neighbors; the exceptions being borderlines, vertices, and facets, which have the same distance between two or more particles. 
In a given three-dimensional distribution of particles, if the volume of the Vorono\"{\i} cells is smaller compared to the cells in neighboring regions, then the particles belong to a clustering region.
%Therefore, the local particle concentration  is inversely proportional to  the volume of the Vorono\"{\i} cells. 
It has been found that a $\Gamma$-distribution can well describe the Probability Density Functions (PDF) of the Vorono\"{\i} volumes of randomly distributed particles in three-dimensions (\cite{Ferenc2007}), namely \vspace{-.1 cm}
%The $\Gamma$-distribution for randomly distributed particles in is described by the following expression from \citet{Ferenc2007}:
%In the 3D case,  the $\Gamma$-distribution has the following pre-factor and exponent \cite[]{Ferenc2007}:
\begin{equation}
\label{Gamma function}
f(x)=\frac{3125}{24}x^4 \exp (-5x),
\vspace{-.15 cm}
\end{equation}
%\begin{equation}
%\label{Gamma function}
%f(x)=\frac{343}{15}\sqrt{\frac{7}{2\pi}}x^{5/2} \exp (-\frac{7}{2}x),
%\end{equation}
where $x = \mathcal{V}/\mathcal{\overline{V}}$ is the Vorono\"{\i} volume $\mathcal{V}$ normalized by the mean volume $\mathcal{\overline{V}}$.
Such a random distribution of particles, and their corresponding Vorono\"{i} cells are shown in the upper panel of figure~\ref{fig:VoronoiExamp}(b).
In the lower panel of figure~\ref{fig:VoronoiExamp}(b), the corresponding $\Gamma$-distribution fitted PDF is shown.
Particles which are not randomly distributed will have a PDF that deviates from this $\Gamma$-distribution. 

Figure~\ref{fig:VoronoiExamp}(a) and \ref{fig:VoronoiExamp}(c) show examples of preferential clustering and grid-like distribution.
In figure~\ref{fig:VoronoiExamp}(a), the particles prefer to aggregate in a small central region, accompanied by void regions. 
We refer to this situation as `preferential clustering'.
In this case, the probabilities of small and large Vorono\"{\i} volumes are higher than the $\Gamma$-distribution. 
In figure~\ref{fig:VoronoiExamp}(c) particles keep the same distance between each other, having the same size Vorono\"{i} cells.
Therefore, the size distribution becomes narrower compared to the case of randomly distributed particles.
We refer to this situation as `grid-like structure'.
% In contrast, probabilities of small and large Vorono\"{\i} volumes are higher than the this $\Gamma$-distribution for grid-like structure. 
%Figure~\ref{fig:VoronoiExamp} also presents the PDFs of the Voronoi Volumes for each cases. 
%As it mentioned above, the randomly distributed case (b) agrees well with the the $\Gamma$-distribution.
 \citet{Tagawa2012} found that these distributions can be well fitted  by a $\Gamma-$distribution with a single fitting parameter $\sigma$, which is the standard deviation of Vorono\"{\i} volumes.
Furthermore, this parameter $\sigma$ can be used for a quantification of the particle clustering.
In this work, we use $\sigma$ to investigate the morphology of the bubble clustering.
Here $\sigma$ is normalized by the standard deviation of randomly distributed particles $\sigma_{rnd}$.
The indicator is (see figure~\ref{fig:VoronoiExamp}): $\sigma$/$\sigma_{rnd} >$ 1 for preferential clustering, $\sigma$/$\sigma_{rnd}$ = 1 for a random distribution, and $\sigma$/$\sigma_{rnd} <$ 1 for a grid-like structure.
%\textcolor{red}{WE WERE HERE on 26th July}

In the application of the Vorono\"{i} analysis on the present numerical data, there are two specific issues that are addressed below.
First, the Vorono\"{\i} cells of particles located near the edges of the domain are not well-defined, i.e. the Vorono\"{i} cells either do not close or close at points outside of the domain. 
These Vorono\"{i} cells located near the domain edges are usually discarded from the Vorono\"{i} analysis (as in \citet{Tagawa2012}).
However, in the present data, the number of bubbles in the domain are small.
In this case, we cannot afford to ignore the edge cells, as it will result in poor statistics.
We take advantage of the periodic boundary condition of the numerics to overcome this problem.
The periodic boundary condition enables us to form a box (3$\times$3$\times$3 larger, and including all particles) surrounding the original box.
We then apply the three-dimensional Vorono\"{i} tesselation on the particle positions in this larger box. % which connected to the interested box, i.e. a box 3$\times$3$\times$3 larger than the original box.
We can now ignore the cell edges on this larger box, as we have sufficient number of particles for good statistics.
Also, if one considers the central box, although some  Vorono\"{i} cells protrude into the neighboring boxes, the total volume is still conserved owing to the periodic boundary conditions.
This is an added advantage of this method.

%\begin{figure}
%	\centering
%\includegraphics[width=.55\textwidth]{NumOfParRandomEffect.eps}
%	\caption{The standard deviation of Vorono\"{i} volumes as a function of the number of randomly distributed point-like particles in a periodic box. The standard deviations are normalized by the standard deviation $\sigma_{\Gamma}$ = 0.4472 for an infinite number of particles in a box, as in ref. \cite{Ferenc2007}. In the present datasets, we consider 16 bubbles in a periodic box, and the corresponding datapoint is indicated using a red triangle marker.}
%	\label{fig:NumberEffect}
%\end{figure}

\begin{figure}
	\centering
\includegraphics[width=1\textwidth]{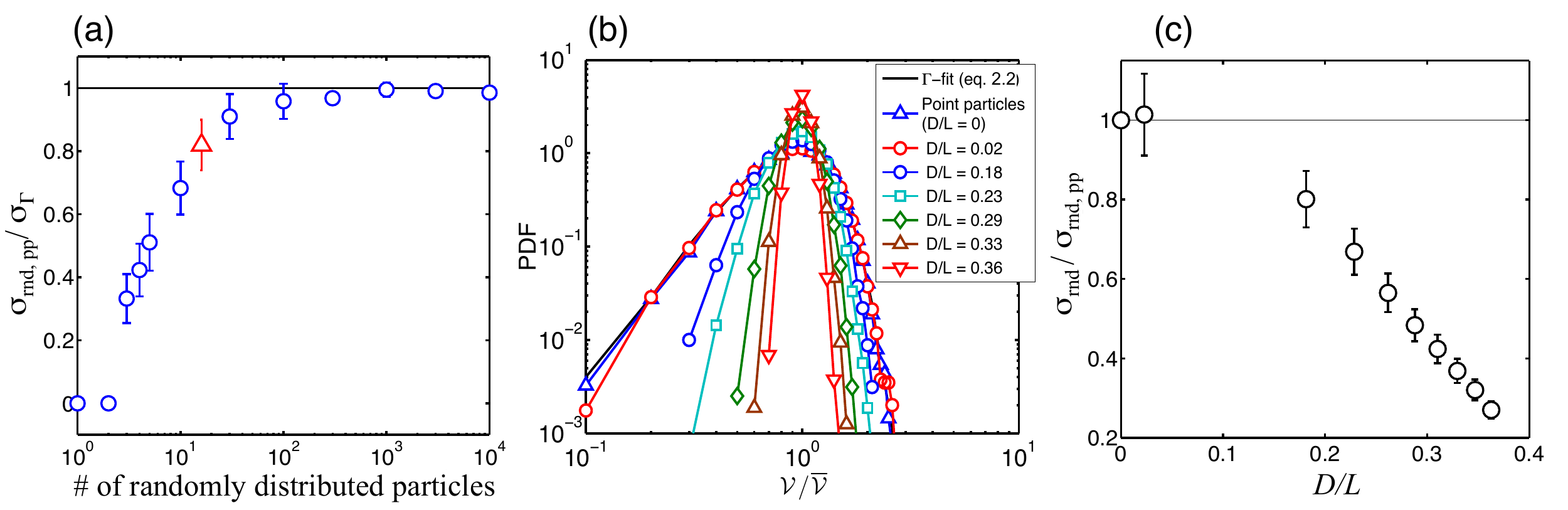}
	\caption{(a) The standard deviation of Vorono\"{i} volumes as a function of the number of randomly distributed point-like particles in a periodic box. The standard deviations are normalized by the standard deviation $\sigma_{\Gamma}$ = 0.4472 for an infinite number of particles ($>$ 10$^6$) in a box, as shown by \cite{Ferenc2007}. In the present datasets, we consider 16 bubbles in a periodic box, and the corresponding datapoint is indicated using a red triangle marker.
	(b) The PDFs of Vorono\"{i} volumes for randomly distributed spheres at different sphere-domain length ratios $D/L$. The $\Gamma$-fit for 16 spheres expressed by equation~(\ref{eq:fitting}) with $\sigma_{rnd, pp}$ (thin black line) and the PDF for randomly distributed point-like particles agree well. The shape of the PDFs becomes narrower with increasing $D/L$ due to the finite-size effect. 
	(c) The standard deviations of Vorono\"{i} volumes $\sigma_{rnd}$ normalized by that for the point-particle case $\sigma_{rnd, pp}$ as a function of $D/L$. The value $\sigma_{rnd}/\sigma_{rnd, pp}$ decreases with increasing $D/L$.}
	\label{fig:RandomFiniteEffect}
	\vspace{-.4 cm}
\end{figure}

Secondly, the number of particles available for the Vorono\"{i} analysis is a key parameter that can significantly affect the results (see Fig.6 in \citet{Tagawa2012}).
We check the dependence of the number of particles on the standard deviation of Vorono\"{i} volumes in figure~\ref{fig:RandomFiniteEffect}(a) for randomly distributed point-like particles.
We vary the number of particles inside each of the boxes that are replicated to form the larger box as mentioned above.
The Vorono\"{i} tesselations are applied and the standard deviation of Vorono\"{i} volumes for each case of the varying number of particles is shown in figure~\ref{fig:RandomFiniteEffect}(a).
Also,  the standard deviation of Vorono\"{i} volumes is normalized by that of randomly distributed particles with numbers $>$10$^6$ (\citet{Ferenc2007}).
Each error bar has been calculated by repeating this procedure more than 10$^4$ times.
We see in figure~\ref{fig:RandomFiniteEffect}(a) that when the particle number is less than 100, the value of the standard deviation changes quite significantly.
%Figure~\ref{fig:NumberEffect} shows the standard deviation for randomly distributed particles as a function of numbers of particles in a periodic box.
We note the peculiarity that when a box includes just one or two particle(s), the Vorono\"{i} volumes are the same or half of the volume of the domain, respectively, and hence the standard deviation is zero.
For a larger number of particles, the standard deviation grows with increasing number of particles and shows an asymptotic behavior, and saturates to the value of unity when the particle numbers approach $\sim$ 1000.
In previous work (\cite{Tagawa2012}) we have used a value of $\sigma/\sigma_{\Gamma}$=1 for the Vorono\"{i} analysis as we had 1000 particles in our simulations.
In the present work, the number of bubbles used in the numerical simulations is 16.
In this case, figure~\ref{fig:RandomFiniteEffect}(a) gives the corresponding value of the standard deviation as $\sigma_{rnd, pp}/\sigma_{\Gamma}$ = 0.82. 
We  account for this change by using $\sigma_{rnd, pp}$ = 0.82$\sigma_{\Gamma}$ in the equation which describes the Vorono\"{i} volume PDF fit using the single parameter $\sigma$ (\cite{Tagawa2012}) :
%This difference can be adopted with applying the corresponding $\sigma$ for following equation.
\vspace{-.15 cm}
\begin{equation}
f(x) = \frac{1}{\sigma^{(2/\sigma^2)}\Gamma(\frac{1}{\sigma^2})}x^{(1/\sigma^2)-1}\exp^{-(x/\sigma^2)}.
\label{eq:fitting}
\vspace{-.15 cm}
\end{equation}
This equation indeed results in a nice fit as shown in figure~\ref{fig:RandomFiniteEffect}(b), where we plot the Vorono\"{i} volume PDF for 16 randomly distributed particles (blue upper triangles) and the curve from equation~(\ref{eq:fitting}) (thin black line). 

All the above discussions were devoted to point-like particles, but in this study we consider bubbles with a finite-size ($D=1-3.5$ mm).
Table~\ref{tab:Cond} lists the different parameters used in the numerics.
%Bubbles in numerics have finite-size and 
Thus, we need to first understand the effect of finite particle size on the Vorono\"{i} volume distributions.
For this, we artificially generate random positions in three-dimensions (see \S~\ref{sec:Simulation})  for 16 perfect spheres of diameter $D$ and change the domain size $L$ to vary the sphere-domain length ratio $D/L$. % the void fraction $\alpha$.
%The method for generating randomly distributed positions is described in 
This sphere-domain length ratio $D/L$ is related to the void fraction $\alpha$ by the expression:
$D/L =  ( 3\alpha/{8\pi})^{1/3}$.
For clarity of presentation, we have chosen to describe the clustering results using $D/L$ instead of $\alpha$.
In figure~\ref{fig:RandomFiniteEffect}(b) we show the Vorono\"{i} volume PDFs for the 16 randomly distributed spheres at different $D/L$. 
%The bubble size compared to the box size is indicated by void fraction $\alpha$.
The PDF of the Vorono\"{i} volume for the randomly distributed point-particles and for spheres at the small value of $D/L$ = 0.02 show quite a similar behavior, i.e. the finite-size effect is then negligible.
The finite-size effects become more significant with increasing $D/L$, and this is seen in the shape of the PDF. The PDFs become narrower with increasing $D/L$, implying that the bubbles are distributed more evenly throughout the domain.
At a large value of $D/L$, each of the spheres occupy relatively larger volumes in the box, which reduces the available free-space (for other spheres), leading to a more constrained distribution and narrower PDF shape.

The standard deviations of the Vorono\"{i} volume PDFs as a function of $D/L$ are shown in figure~\ref{fig:RandomFiniteEffect}(c).
The values $\sigma_{rnd}$ are normalized by the standard deviation obtained from the $\Gamma$-distribution fit for randomly distributed point particles, $\sigma_{rnd, pp}$.
The indicator $\sigma_{rnd}/\sigma_{rnd, pp}$ decreases monotonically with $D/L$, starting at 1 (at $D/L$ = 0) and reduces to $\sim$1/5 for $D/L = 0.36$, clearly indicating the effect of finite-size.
The normalization of the clustering indicator $\mathcal{C} = \sigma(D/L)/\sigma_{rnd}(D/L)$ for each case used in the discussion below is carried out at the same bubble-domain length ratio $D/L$, in order to fully focus on dynamical effects.
The total runtime for numerical simulations is about t=2 s, and the Vorono\"{i} volume time series reveals that the clustering is initially transient and settles to a quasi-steady state after t=1 s.
Hence, we only consider data after t=1 s from the starting time.
The clustering results are averaged over different snapshots at intervals of t=0.05 s. 
\vspace{-.5 cm}
\section{Numerical method}
\label{sec:Simulation}
%\subsection{Governing equations}

Three-dimensional direct numerical simulations (DNS) have been performed to simulate bubbles rising in a swarm, using periodic boundary conditions in all directions to mimic an `infinite' swarm without wall effects, similarly to what has been done by \cite{Bunner2002}. The simulations have been carried out using a model that incorporates the front-tracking (FT) method (\cite{Unverdi1992}), which tracks the interfaces of the bubbles explicitly using Lagrangian control points distributed homogeneously over the interface. Compared to interface reconstruction techniques, such as volume-of-fluid or level-set methods, the advantage of the FT method is that the bubbles are able to approach each other closely (within less than the size of 1 grid cell) and can even collide, while preventing (artificial) merging of the interfaces. Therefore, the size of the bubbles remains constant throughout the simulation. Especially for bubble swarm simulations with high void fractions as studied in this work, this is an important aspect. In addition, the interface is sharp allowing the surface tension force to act at the exact position of the interface. 

In our model (see \cite{Dijkhuizen2010-model,Roghair2011a} for details), the fluid flow is solved by the discretised incompressible Navier-Stokes equations on a Eulerian background mesh consisting of cubic computational cells:
\begin{equation}
%\begin{align}
 \rho \frac{\partial \vec{u}}{\partial t} + \rho\nabla\cdot\left( \vec{u}\vec{u}\right) = -\nabla p  + \rho\vec{g} +\nabla \cdot \mu\left[\nabla \vec{u} + (\nabla \vec{u})^T \right]+ \vec{F}_\gamma  ,~~~~~~~
  \nabla\cdot\vec{u}=0 
  \label{eq:divu}
%\end{align}
\end{equation}
where $\vec{u}$ is the fluid velocity and $\vec{F}_\gamma$ representing a singular source-term accounting for the surface tension force at the interface (see below). The flow field of both phases is resolved using a one-fluid formulation where the physical properties are determined from the local phase fraction; the local density $\rho$ is obtained by the weighted arithmetic mean and the dynamic viscosity $\mu$ is obtained via the weighted harmonic mean of the kinematic viscosities. %at position $\vec{x}$) vary according to the phase fraction $\phi$ in each cell:
%
%\begin{subequations}
% \begin{align}
 %\rho\left( \vec{x} \right) &= \sum_{p=0}^{n_\text{phase}-1}\phi_p\left(\vec{x} \right)\rho_p \label{eq:macrho} \\
  %\frac{\rho\left( \vec{x} \right)}{\mu\left( \vec{x} \right)} &= \sum_{p=0}^{n_\text{phase}-1}\phi_p\left(\vec{x} \right)\frac{\rho_p}{\mu_p} \label{eq:maceta} 
 %\end{align}
%\end{subequations}
%
The interface between the gas and the liquid is tracked using Lagrangian control points, distributed over the interface. The control points are connected such that they form a mesh of triangular cells. The surface tension force is acquired by obtaining the pull-forces for each marker $m$ and its neighboring cells $i$: $\vec{F}_{\gamma, i\rightarrow m} =\gamma\left(\vec{t}_{mi} \times \vec{n}_{mi} \right)$.
%
% \begin{subequations}
%\begin{equation}
%\label{eq:surftens}
%\vec{F}_{\gamma, i\rightarrow m} =\gamma\left(\vec{t}_{mi} \times \vec{n}_{mi} \right)
%\end{equation}
%
The shared tangent $\vec{t}_{mi}$ is known from the control point locations, and the shared normal vector $\vec{n}_{mi}$ is obtained by averaging the normals of marker $m$ and neighboring marker $i$. 
% As one term can be discarded due to orthogonality, the resulting surface tension on a marker $m$ is obtained as:
%
% \begin{equation}
%  \vec{F}_{\gamma,m} =  \frac{1}{2}\left( \underbrace{ ( \vec{t}_{mi} \times \vec{n}_{m} ) }_{=0} + ( \vec{t}_{mi} \times \vec{n}_{i}) \right) = \tfrac{1}{2}\gamma \sum_{i=a,b,c} \left( \vec{t}_{mi} \times \vec{n}_{i} \right)
% \end{equation}
% \end{subequations}
%
Subsequently, the surface tension force is mapped to the Eulerian background grid using mass-weighing \citep{Deen2004} at the position of the interface. After accounting for the surface tension force on all interface cells, the total pressure jump $\Delta p$ of the bubble is obtained. The pressure jump is distributed over the bubble interface and mapped back to the Eulerian mesh. For interfaces with a constant curvature (i.e. spheres), the pressure jump and surface tension cancel each other out exactly on each marker, but if the curvature varies over the interface (which is the case for deformed bubbles), a small net force will be transmitted. 
%More details of this procedure are given in \cite{Dijkhuizen2010-model}.

At each time step, after solving the fluid flow equations, the Lagrangian control points are advected with the interpolated flow velocity. Spatial interpolation of the flow field to the control point positions is performed by a piecewise cubic spline, and temporal integration is performed by Runge-Kutta time stepping. Since the control points may move away from or towards each other, the interface mesh is remeshed afterwards, in order to keep the control points equally distributed on the interface (while keeping the volume enclosed by the dispersed element constants).

Each bubble was tracked individually, using the locations of the control points on the interface to acquire the bubble position (center of mass) and bubble shape (aspect ratio), which were stored for further analysis. %\subsection{Numerical method}
% Since the bubbles are deformable, the bubble size is represented as the equivalent bubble diameter $d_{eq}$ or $d_b$ obtained using the bubble volume $V_b$:
% %
% \begin{equation}
% \label{eq:deq}
%  d_{eq} = \sqrt[3]{\frac{6V_b}{\pi}}
% \end{equation}
% % 
The aspect ratio is calculated from the ratio between the major and the minor axis $\chi$ along the Cartesian axes:  $\chi = ({\sqrt{d_xd_y}})/{d_z}$.
%
%\begin{equation}
% \label{eq:aspect}
% \chi = \frac{\text{major axis}}{\text{minor axis}} = \frac{\sqrt{d_xd_y}}{d_z}
%\end{equation}
Note that this procedure neglects diagonal shape deformations, so that strongly deformed bubbles oriented diagonally may be attributed an aspect ratio of $\chi \approx 1$ (nearly spherical). 
In this work we consider bubbles with limited deformability and under mild flow conditions so that these effects can safely be neglected.

%Such deformations have been seen to occur in simulations with highly deformable bubbles, such as $6.0$ \milli\meter\ air bubbles in water, in which case it would be benificial to use a different method of calculating the aspect ratio, but 
%Note that this equation neglects the diagonal shape deformations, so that strongly deformed bubbles oriented diagonally may be attributed an aspect ratio of $\chi \approx 1$ (nearly spherical). Such deformations have been seen to occur in simulations with highly deformable bubbles, such as $6.0$ \milli\meter\ air bubbles in water, in which case it would be benificial to use a different method of calculating the aspect ratio, but in this work we work with bubbles with limited deformability. As a check, we have run a simulation with $4.0$ \milli\meter\ air bubbles in water, and the amount of bubbles deformed along the Cartesian axes strongly outnumbers the bubbles deformed diagonally, and these effects can safely be neglected.

%\tr{(From DETLEF)So you miss/neglect ``diagonal'' bubble effects? Comment on this. For turbulent flow they will play a role. (IVO: please comment on this issue)}

%\subsection{Simulation settings}

\begin{table}
\vspace{-.6 cm}
\caption{Summary of the simulation parameters.} 
\vspace{-.15 cm}
%$V_{mean}$: water mean flow speed, $\mathrm{Re}_{\lambda}=(15u_{rms}^4/\epsilon\nu)^{1/2}$: Taylor-Reynolds number, $u_{rms}$: mean velocity fluctuation, $\eta=(\nu^3/\epsilon)^{1/4}$ and $\tau_{\eta}=(\nu/\epsilon)^{1/2}$: are the Kolmogorov's length scale and time scale respectively, $L$: integral length scale of the flow, $\epsilon$: mean energy dissipation rate,  $St=\tau_p/\tau_{\eta}$: Stokes number,  and $N_{data}$: number of data points used to calculate the Lagrangian statistics.}
\begin{center}
 \begin{tabular}{|c|c|c|c|c|c|}
 \hline 
       Case&Diameter &Void fraction& Sphere-domain ratio & Kinematic viscosity & Surface tension \\ \vspace{-.15 cm}
       $\#$&$D$ [mm]&$\alpha$ [$\%]$& $D/L$ & $\nu$ [$\times$10$^{-6}$ m$^2$/s]& $\gamma$ [mN/m] \\ \hline 
           1-3   & 1.0 & 10, 25, 40 & 0.23, 0.31 ,0.36  & 1  & 73 \\ 
           4      & 1.0 & 10  & 0.23   & 5 & 73 \\ 
           5      & 1.0 & 10  & 0.23  & 1  & 7.3\\
           6-11 & 2.0 & 5 - 40 & 0.18 - 0.36  & 1 & 73 \\ 
           12-18 & 2.5 & 5 - 40 & 0.18 - 0.36  & 1 & 73 \\ 
           19-26 & 3.0 & 5 - 40 & 0.18 - 0.36  & 1 & 73 \\ 
           27-33 & 3.5 & 5 - 40 & 0.18 - 0.36  & 1 & 73 \\           
           \hline        
    \end{tabular}
    \end{center} \label{tab:Cond}
    \vspace{-.5 cm}
\end{table}

Simulations were performed using 16 bubbles in a periodic domain. For ellipsoidal bubbles, \cite{Bunner2002} have indicated that 12 bubbles is the minimum number of bubbles that are required to simulate bubbles rising in a swarm, based on their terminal rise velocity. The void fraction $\alpha=V_{bubbles}/V_{domain}$ was varied from dilute ($\alpha=0.05$) up to dense void fractions of $\alpha=0.4$ by changing the domain size. In all simulations, the spatial resolution was determined by the bubble diameter $1.0\cdot10^{-3}\leq d_b\leq3.5\cdot10^{-3}$ such that the length of a cubic grid cell $\Delta x = d_b/20$. The physical properties represent typical air bubbles in water conditions, i.e. a density ratio of $\frac{\rho_\text{liquid}}{\rho_\text{gas}}\approx 1000$, dynamic viscosity ratio $\frac{\mu_\text{liquid}}{\mu_\text{gas}}\approx 50$ and a surface tension coefficient $\gamma$= 0.073 \newton\per\meter. From the simulations, an initial transient time of 0.2 \second\ is discarded.
The simulation parameters are summarized in table~\ref{tab:Cond}. 

%\subsection{Randomization procedure}
%\label{sec:RandProc}
An initial structured configuration of the bubble positions can cause streaming (\cite{Bunner2003}) and liquid channeling, and may take a significant part of the simulation time to break up into a non-structured configuration. To prevent such initialization effects from influencing our simulation results, the bubble positions were initially set to a non-ordered fashion in the domain. Especially at higher void fractions it is not efficient to subsequently place a bubble randomly in the domain without allowing overlap. Therefore, a Monte-Carlo simulation procedure has been used to generate the initial positions of the bubbles which works for all void fractions (\cite{Frenkel2002, Beetstra2005}). 
First, the bubbles are placed as spheres in a structured configuration in the domain in a simple cubic configuration. Depending on the void fraction, the bubbles might overlap with each other. We now define the potential energy of the system as: $E=\left[| \vec{x}_i - \vec{x}_j|/(R_i+R_j) \right]^n$, with a variable $n$ characterizing how steep the potential is.
%
%\begin{equation}
% \label{eq:potentialenergy}
% E=\left(\frac{\left| \vec{x}_i - \vec{x}_j\right|}{R_i+R_j} \right)^n,
%\end{equation}
%
The position of a bubble $i$ is given by $\vec{x}_i$ and its radius by $R_i$. Each bubble is now moved by a small amount in a random direction and the potential energy is determined. A move is accepted whenever the potential energy remains equal or becomes less, whereas a move that increases the potential energy is accepted only if it is smaller than a critical number $c$: $c=\exp[{k(E_\text{old}-E_\text{new})}]$,
where $k$ was set to 50. In a single iteration, each bubble is allowed 200 (attempts of) displacement. Then, the power $n$ is gradually increased from an initial value of 6 up to a final value of 100, and a new iteration starts. The potential energy of the system as a whole decreases during this process, and when the final state has been reached and the bubbles show no overlap at all, the positions of the bubbles are accepted for use as starting positions in the front tracking model. Additionally, in the initial transient of the front tracking simulations, the bubbles accelerate, deform and move through the domain, which also changes their relative positions. This start-up stage is discarded from further analysis.
For the random positions of the point-particles as shown in figure~\ref{fig:RandomFiniteEffect}(c), the same procedure was used, except that we chose the number of allowed displacements for each particle per iteration to be 10$^4$.
\vspace{-.5 cm}

%%%---------------------------------------- RESULTS ------------------------------------------------%%%
%%%---------------------------------------- RESULTS ------------------------------------------------%%%
%%%---------------------------------------- RESULTS ------------------------------------------------%%%

\section{Results and Discussions}
\label{sec:RAD}

\begin{figure}
        \centering
	\includegraphics[width=.85\textwidth]{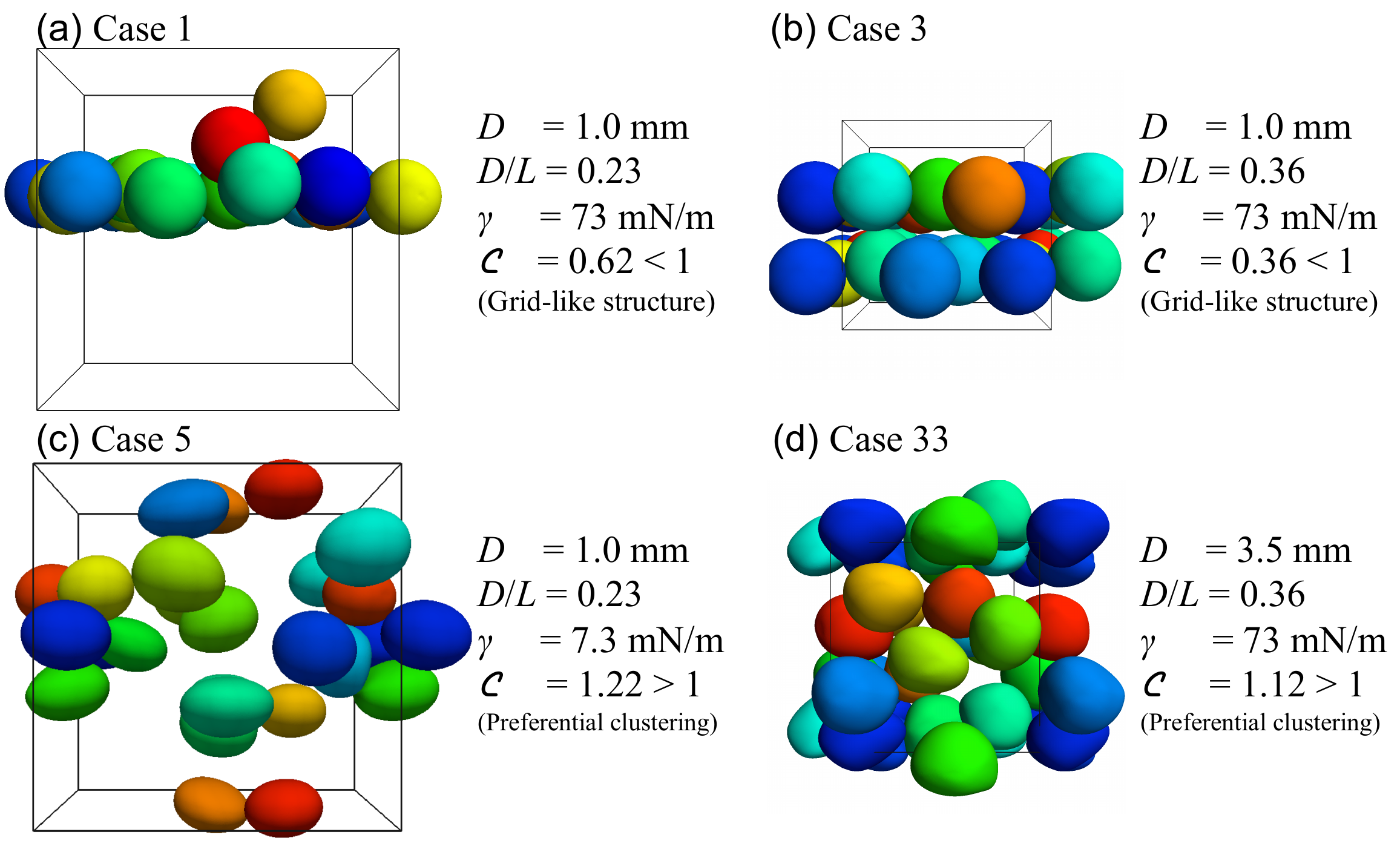}
        \caption{Snapshots of bubbles showing typical clustering morphologies. The periodic box is indicated by the thin black lines. The bubbles are colored to aid readers to distinguish between individual bubbles.}
        \label{fig:Snapshot}
        \vspace{-.4 cm}
\end{figure}

We recall that the value of the clustering indicator $\mathcal{C} = \sigma$/$\sigma_{rnd}$ are $\mathcal{C}>1$ for preferential clustering, $\mathcal{C} = 1$ for a random distribution, and $\mathcal{C} < 1$ for a grid-like structure (figure~\ref {fig:VoronoiExamp}). 
\\
In figure~\ref{fig:Snapshot}, we now show the typical bubble clustering snapshots.
Figure~\ref{fig:Snapshot}(a) displays a snapshot of the case of bubble diameter $D$ = 1.0 mm at $D/L$ = 0.23. 
We observe horizontal clustering in one layer. 
This snapshot corresponds to $\mathcal{C} < 1$, a grid-like cluster morphology. 
In figure~\ref{fig:Snapshot}(b), $D$ = 1.0 mm at $D/L$ = 0.36, and the bubbles show horizontal clustering in a double layer, owing to larger $D/L$ (i.e. larger void fraction $\alpha$), and the value of $\mathcal{C}$ is correspondingly less than 1. 
It must be noted that the shapes of the bubbles for $D$ = 1.0 mm at (a) $D/L$ = 0.23 and (b) $D/L$ = 0.36 are almost spherical. 
In figure~\ref{fig:Snapshot}(c), we show a case of $D$ = 1.0 mm at $D/L$ = 0.23 with lower surface tension, where the bubbles are evenly distributed and the horizontal one-layer clustering no longer prevails. 
In figure~\ref{fig:Snapshot}(d), in the case of $D$ = 3.5 mm at $D/L$ = 0.36, the bubbles are more evenly distributed. 
The corresponding $\mathcal{C}$ values for both cases are larger than 1, i.e. indicating preferential clustering.
%Figure~\ref{fig:AspectNormSigma}(b) indeed indicates preferential clustering as discussed before. 
The bubbles with $D$ = 1.0 mm at $D/L$ = 0.23 and those with $D$ = 3.5 mm at $D/L$ = 0.36 have a deformed shape. %, and hence show a preferential clustering structure due to the dynamical flow effects. 

\begin{figure}
	\centering
  \includegraphics[width=1\textwidth]{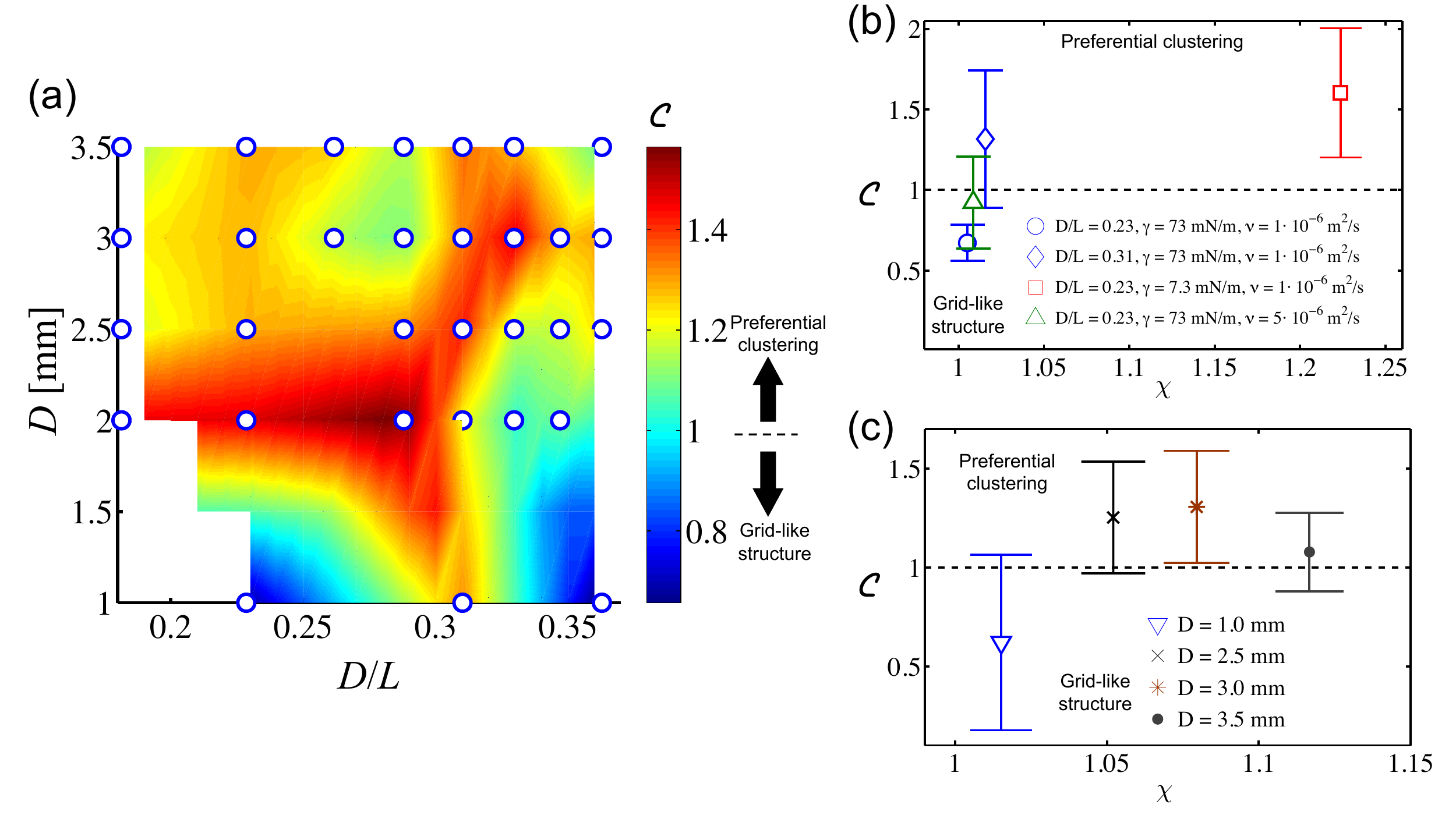}
%  \subfloat{	\includegraphics[width=.75\textwidth]{Chi_NormSigma.pdf}}
%\subfloat[The average aspect ratio $\chi$ vs. the void fraction $\alpha$.]{\label{fig:AspectAll}\includegraphics[width=.45\textwidth]{Alpha_chi.eps}}
	\caption{\textit{Left panel}: (a) Bubble clustering results: A contour plot of the clustering indicator $\mathcal{C}$ as function of the bubble-domain length ratio $D/L$ for all bubble sizes $D$ (in mm). The colorbar indicates the magnitude of $\mathcal{C}$. All the simulation cases in table~\ref{tab:Cond} are shown here  except 4 and 5.
	\textit{Right panel}: The normalized standard deviation of the Vorono\"{i} volumes $\mathcal{C}$, as a function of the aspect ratio $\chi$ for: (b) fixed bubble size, $D$ = 1.0 mm (cases 1, 2, 4, 5 in table~\ref{tab:Cond}), and (c) fixed bubble-domain length ratio, $D/L$ = 0.36 (cases 3, 18, 26, 33 in table~\ref{tab:Cond}). The standard deviation increases with increasing aspect ratio, indicating that the shape of the bubbles plays a crucial role in determining the  clustering morphology. Spherical bubbles with an aspect ratio $\chi \lesssim$ 1.015 have $\mathcal{C} <1$, indicating the grid-like structure. All deformed bubbles with $\chi \gtrsim$ 1.015 have $\mathcal{C} >1$, implying preferential clustering.
	%The normalized standard deviation values must be interpreted as shown in figure~\ref {fig:VoronoiExamp}.
	}
\label{fig:NormSigma}
\vspace{-.3 cm}
\end{figure}

For a quantitative discussion, in figure~\ref{fig:NormSigma}(a) we show (as a color contour plot) the values of the clustering indicator $\mathcal{C}$ at different bubble-domain length ratios $D/L$ and bubble sizes $D$. 
The data points in the parameter space (table~\ref{tab:Cond}) are indicated using open blue circles.
%The normalized standard deviation is shown by color scale.
The formation of grid-like clusters ($\mathcal{C}<$1) were only encountered in the cases of 1.0 mm diameter bubbles at $D/L$ = 0.23 and 0.36 (figure~\ref{fig:NormSigma}(a)). All other cases show preferential clustering ($\mathcal{C}>$1), but to a different extent.

It is well-known that a rising spherical bubble with a free-slip boundary condition generates little vorticity, whereas a rising deformed bubble has a wide region of wake structure behind it (\cite{Magnaudet2007}). The amount of vorticity generated from the bubbles determines the clustering morphology. The flow around spherical bubbles can be expected to be close to potential flow, containing little vorticity, and these bubbles form a grid-like cluster (in the horizontal plane). Deformable bubbles with large wake regions show aggregation in the vertical direction. Hence, in the discussion of the results, we focus on the bubble shape (or deformability) characterized by the bubble aspect ratio $\chi$. Below, we fix the size and bubble-domain length ratio, and discuss the clustering at 
different $\chi$.

\begin{comment}
\begin{figure}
	\centering
	\includegraphics[width=1\textwidth]{Chi_NormSigma.pdf}
%	\subfloat[Bubbles of $D$ = 1.0 mm]{\includegraphics[width=.4\textwidth]{Chi_NormSigma_d1mm.eps}}
%	\subfloat[Bubbles with $D/L$ = 0.36]{\includegraphics[width=.4\textwidth]{Chi_NormSigma_Alpha40pct.eps}}
	\caption{The normalized standard deviation of the Vorono\"{i} volumes $\mathcal{C}$, as a function of the aspect ratio $\chi$ for: (a) fixed bubble size, $D$ = 1.0 mm (cases 1, 2, 4, 5 in table~\ref{tab:Cond}), and (b) fixed bubble-domain length ratio, $D/L$ = 0.36 (cases 3, 18, 26, 33 in table~\ref{tab:Cond}). The standard deviation increases with increasing aspect ratio, indicating that the shape of the bubbles plays a crucial role in determining the  clustering morphology. Spherical bubbles with an aspect ratio $\chi \lesssim$ 1.015 have $\mathcal{C} <1$, indicating the grid-like structure. All deformed bubbles with $\chi \gtrsim$ 1.015 have $\mathcal{C} >1$, implying preferential clustering.} 
	\label{fig:AspectNormSigma}
\end{figure}
\end{comment}

First, we keep the bubble size constant ($D$ = 1.0 mm) and discuss results for different bubble-domain length ratios, surface tension, and viscosity. Figure~\ref{fig:NormSigma}(b) shows the values of the clustering indicator $\mathcal{C}$ as a function of the bubble aspect ratio $\chi$. The value of $\mathcal{C}$ increases with an increase in the aspect ratio, indicating that the bubble shape is crucial for the clustering structure. 
We now fix the bubble-domain length ratio ($D/L$ = 0.36) and study the clustering behavior for different bubble sizes in figure~\ref{fig:NormSigma}(c).
%, we again show the variation of $\mathcal{C}$ at different aspect ratios $\chi$. 
Although the fixed parameter is different in this case, we still see the same trend of increasing $\mathcal{C}$ with increasing aspect ratio $\chi$.
Overall, we find that the shape of the bubbles is crucial for the structure of the clustering. Spherical bubbles with $\chi \lesssim$ 1.015 have $\mathcal{C} <1$, indicating the grid-like clustering structure. All the deformed bubbles with $\chi \gtrsim$ 1.015, have $\mathcal{C} >1$, indicating the preferential clustering morphology.

%\textcolor{red}{(1) We must compare our results to others in literature, who have used pair correlation or other methods, (2) Also, some more discussion of the physics - vorticity generation from bubbles (citing more papers) -  this will add to the text, and increase discussion} 

%Overall, we observe that spherical bubbles with $\chi \le$ 1.015 give $\sigma/\sigma_{rnd} <1$, indicating the grid-like structure while bubbles of $\chi >$ 1.015 have $\sigma/\sigma_{rnd} <1$, meaning the preferential clustering.

\begin{figure}
\centering
\includegraphics[width=.5\textwidth]{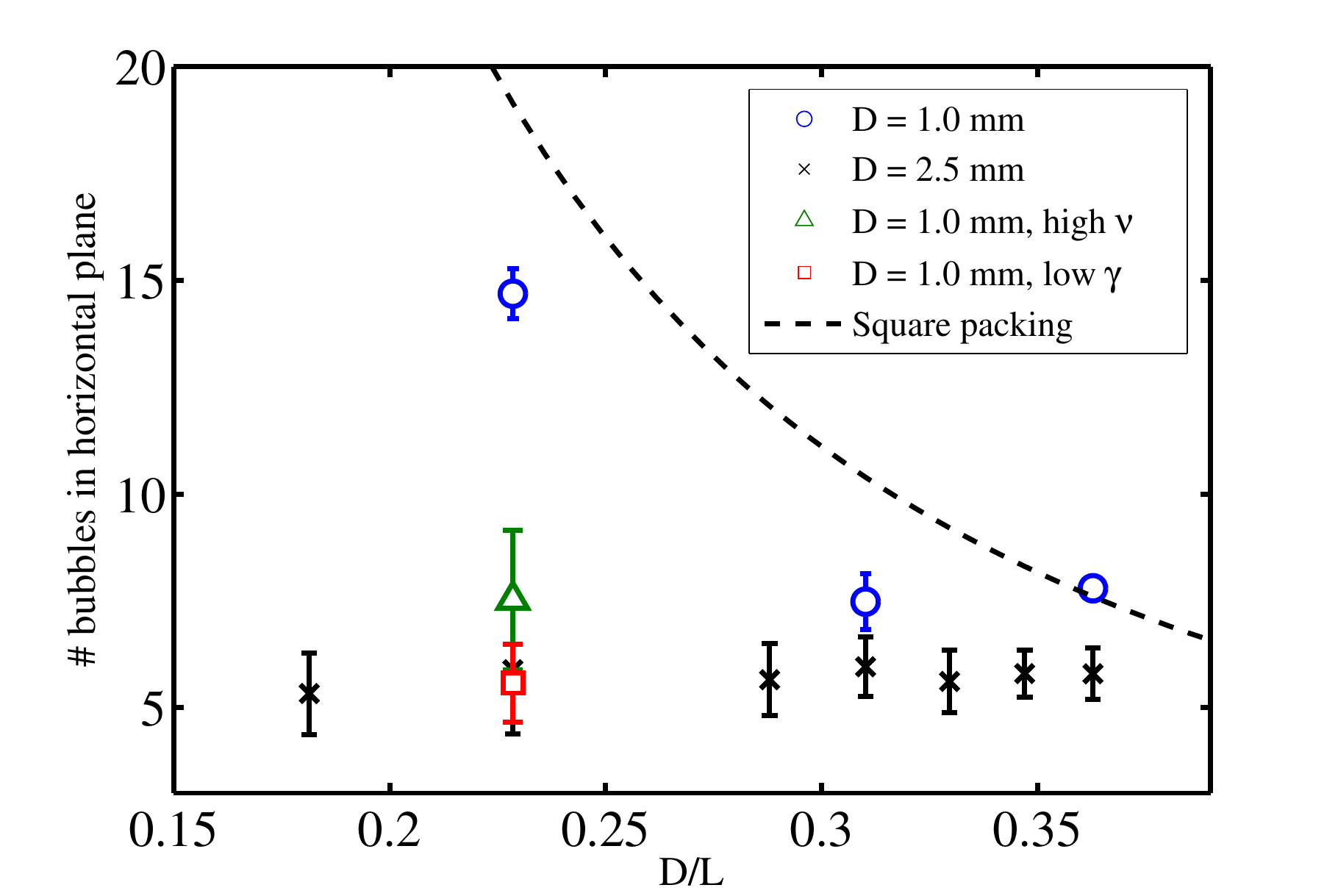}
\caption{The maximum number of bubbles in a horizontal plane. The dotted line shows the values for square packing. The 1 mm bubbles (cases 1-5 in table~\ref{tab:Cond}) show a increasing trend with decreasing $D/L$ indicating that they form horizontal clusters. The 2.5mm bubbles (cases 12-18 in table~\ref{tab:Cond}) show almost constant values of 6, implying that there is no horizontal clustering. The result in the case of liquid with high viscosity (case 4 in table~\ref{tab:Cond}, green triangle) and that for low surface tension (case 5 in table~\ref{tab:Cond}, red square) are also shown.}
\label{fig:HorizontalClustering}
\vspace{-.4 cm}
\end{figure}

We also quantify the intensity of horizontal clustering by counting the maximum number of bubbles at a horizontal plane, for the cases of $D$ = 1 mm and 2.5 mm bubbles. The bubbles are sliced at the horizontal plane and divided by the area of a circle based on the bubble radius. Figure~\ref{fig:HorizontalClustering} shows the maximum number of bubbles in a horizontal plane versus the bubble-domain length ratio. The line obtained from the theory of square packing is also shown for the sake of comparison. On one hand, the 1 mm bubbles show a trend similar to the theoretical line, indicating that they organize themselves to form horizontal clusters. On the other hand, the 2.5 mm bubbles show almost constant values around 6, indicating the absence of horizontal clustering. The result of the lower surface tension case (case 5 in table~\ref{tab:Cond}) is rather close to the 2.5 mm case (case 13 in table~\ref{tab:Cond}) due to deformation. These results for the horizontal clustering are consistent with those obtained from the Vorono\"{i} analysis.
\vspace{-.6 cm}

\section{Conclusion}
\label{sec:Concl}
%\textcolor{red}{ to be written at the end, for now this is the abstract:}\\
In this work, we have applied the three-dimensional Vorono\"i analysis on DNS data of freely rising deformable bubbles in order to investigate the clustering morphology.  
The numerics used a front-tracking method which allows the simulation of fully deformable interfaces of the bubbles at different diameters, bubble-domain length ratios, surface tension, and liquid viscosity. 
The present Vorono\"i analysis takes into account the number of bubbles and finite-size effects.
It then provides a clustering indicator $\mathcal{C} =  \sigma/\sigma_{rnd}$, where $\sigma$ is the standard deviation of Vorono\"i volumes of the bubbles and $\sigma_{rnd}$ is the standard deviation of Vorono\"i volumes of randomly distributed particles.
We quantitatively identify two different clustering morphologies: $\mathcal{C} >$ 1 for preferential clustering and $\mathcal{C}<$ 1 for a grid-like structure.
Our results indicate that the bubble deformability, represented by its aspect ratio $\chi$, plays the most crucial role in determining the clustering morphology. 
%We mainly discern two different types of clustering morphologies: preferential clustering and grid-like clustering.
The grid-like morphology is observed in the case of nearly spherical bubbles with $\chi \lesssim$ 1.015. When the bubbles are deformable, for $\chi \gtrsim$ 1.015, a preferential clustering behavior is observed. 
This clustering behavior is believed to be related to the amount of vorticity generated by the bubbles. 
The preferential clustering for deformed bubbles is due to the low pressure regions in their wakes, which attract other bubbles. 
Spherical bubbles tend to form a grid-like structure due to reduced vorticity generation.

%%%---------------------------------------- ACKNOWLEDGEMENTS ------------------------------------------------%%%
%%%---------------------------------------- ACKNOWLEDGEMENTS ------------------------------------------------%%%
%%%---------------------------------------- ACKNOWLEDGEMENTS ------------------------------------------------%%%
%\section*{Acknowledgments}
We thank L. van Wijngaarden  for fruitful discussions. We acknowledge support from the EU COST Action MP0806 on ``Particles in Turbulence''. We acknowledge support from the Foundation for Fundamental Research on Matter (FOM) through the FOM-IPP Industrial Partnership Program: Fundamentals of heterogeneous bubbly flows.
\vspace{-.5 cm}
%\begin{figure}
%	\centering
%	%\subfloat[Bubbles of 1.0 mm in diameter at $\alpha$ = 10 \%]{\includegraphics[width=.45\textwidth]{Chi_NormSigma_all2_2.eps}}
%	\subfloat[All data]{\includegraphics[width=.45\textwidth]{Chi_NormSigma_all3.eps}}
%	\caption{The normalized standard deviation of the Vorono\"{i} Volumes as a function of the aspect ratio $\chi$. The standard deviation decreases with decreasing aspect ratio, indicating that the shape of the bubbles is crucial for the structure of the clustering. Spherical bubbles of $\chi \ge$ 0.985 give $\sigma/\sigma_{rnd} <1$, indicating the grid-like structure. All deformed bubbles of $\chi <$ 0.985 have $\sigma/\sigma_{rnd} <1$, meaning the preferential clustering. Symbols corresponds to bubbles of 1.0 mm in diameter (blue circles), 2.0 mm in diameter (black star), 2.5 mm in diameter (black cross), 3.0 mm in diameter (black downward-pointing triangle), 3.5 mm in diameter (black diamond), 1.0 mm in high viscosity of surrounding liquid (green triangle), and 1.0 mm with low surface tension (red square).}
%	\label{fig:NormSigma}
%\end{figure}

\bibliographystyle{jfm.bst}

\end{document}